\documentclass{appolb}
\usepackage{graphicx}

\begin{document}

\title{Hadron production within PHSD
\thanks{XI Workshop on Particle Correlations and Femtoscopy 2015, Warsaw, Poland}}

\author{
	P. Moreau$^{1}$, W. Cassing$^{2}$, A. Palmese$^{2}$, E. L. Bratkovskaya$^{1}$
\address{$^{1}$Frankfurt Institute for Advanced Studies and Institut f\"{u}r Theoretische Physik, Johann Wolfgang Goethe Universit\"{a}t,
	Frankfurt am Main, Germany}
\address{$^{2}$Institut f\"{u}r Theoretische Physik, Universit\"{a}t Gie$\ss$en, Germany}
}

\maketitle

\begin{abstract}
	We study the production of  hadrons in nucleus-nucleus collisions 
	within  the Parton-Hadron-String Dynamics (PHSD) transport
	approach that is extended to incorporate essentials aspects of
	chiral symmetry restoration (CSR) in the hadronic sector (via the
	Schwinger mechanism) on top of the deconfinement phase transition as
	implemented in PHSD. The
	essential impact of CSR is found in the Schwinger mechanism (for
	string decay) which fixes the ratio of strange to light quark
	production in the hadronic medium. Our studies provide a microscopic explanation for the maximum in the
	$K^+/\pi^+$ ratio at about 30 A GeV which only shows up if in
	addition to CSR a deconfinement transition to partonic
	degrees-of-freedom is incorporated in the reaction dynamics.
\end{abstract}


\section{Introduction}
\label{intro} In this contribution we summarize the results from our study in Ref. \cite{Cassing:2015owa} that investigates the strangeness enhancement in nucleus-nucleus
collisions \cite{rafelski,stock} or the 'horn' in the $K^+/\pi^+$
ratio \cite{GG99,GGS11}. Previously both phenomena have been addressed to a deconfinement transition. Indeed, the actual
experimental observation could not be described within conventional
hadronic transport theory \cite{Jgeiss,Brat04,Weber} and remained a major
challenge for microscopic approaches.

\pagebreak[4]

\section{Extensions in PHSD3.3}
Our studies are performed within the PHSD transport approach that
has been described in Refs. \cite{PHSD,PHSDrhic}. PHSD incorporates
explicit partonic degrees-of-freedom in terms of strongly
interacting quasiparticles (quarks and gluons) in line with an
equation-of-state from lattice QCD (lQCD) as well as dynamical
hadronization and hadronic elastic and inelastic collisions in the
final reaction phase. 

\subsection{Strings in (P)HSD}
We recall that in the PHSD/HSD, the high energy inelastic hadron-hadron
collisions in the hadronic phase are described by the FRITIOF model \cite{FRITIOF}, where
two incoming nucleons emerge the reaction as two excited color
singlet states, i.e. 'strings'. The production probability $P$ of massive $s\bar{s}$ or
$qq\bar{q}\bar{q}$ pairs is suppressed in comparison to light flavor
production ($u\bar{u}$, $d\bar{d}$) according to the Schwinger-like
formula \cite{Schwinger}, i.e.
\begin{equation}
	{P(s\bar{s}) \over P(u\bar{u})} ={P(s\bar{s}) \over P(d\bar{d})} = \gamma_s = \exp\left(-\pi
	{m_s^2-m_q^2\over 2\kappa} \right) ,
	\label{schwinger}
\end{equation}
with $\kappa\approx 0.176$~GeV$^2$ denoting the string tension
and $m_s, m_q=m_u=m_d$ the appropriate (dressed) strange and light quark masses. 
Inserting the constituent (dressed) quark masses $m_u \approx 0.33$~GeV
and $m_s \approx 0.5$ GeV in the vacuum a value of $\gamma_s \approx 0.3$ is
obtained from Eq.(1). This ratio is expected to be different in a nuclear medium and actually should depend on the in-medium quark condensate $<\bar{q}q>$.

\subsection{The scalar quark condensate}
As it is well known the scalar quark condensate $<\bar{q}q>$ is viewed as an order
parameter for the restoration of chiral symmetry at high baryon
density and temperature. A reasonable estimate for the quark scalar condensate in dynamical
calculations has been suggested by Friman et al. \cite{Toneev98},
\begin{equation}
	\frac{<\bar{q}q>}{<\bar{q}q>_V} = 1 - \frac{\Sigma_\pi}{f_\pi^2
		m_\pi^2}\rho_S - \sum\limits_h{\sigma_h \rho_S^h \over f_\pi^2
		m_\pi^2}, \label{condens2} \end{equation} 
where $\sigma_h$ denotes the $\sigma$-commutator of the relevant mesons $h$ and $\rho_S$ the scalar nucleon density. 
Furthermore, $<\bar{q}q>_V$ denotes the vacuum condensate, $\Sigma_\pi \approx$
45 MeV is the pion-nucleon $\Sigma$-term, $f_\pi$ and $m_\pi$ the
pion decay constant and pion mass, respectively. 

The basic assumption now is
that the strange and light quark masses in the hadronic medium drop
both in line with the ratio (\ref{condens2}),
\begin{equation}   \label{mss}
	m_s^* = m_s^0 + (m_s^v-m_s^0)\left| \frac{<\bar{q}q>}{<\bar{q}q>_V}
	\right| ,  m_q^* = m_q^0 + (m_q^v-m_q^0) \left| \frac{<\bar{q}q>}{<\bar{q}q>_V}
	\right| ,
\end{equation} using $ m_s^0 \approx$ 100 MeV and $m_q^0 \approx 7$
MeV for the bare quark masses while the vacuum (dressed) masses are
$ m_s^v \approx$ 500 MeV and $m_q^v \approx 330$ MeV, respectively.

\section{Comparison of PHSD3.3 results to A+A data}
\label{sec:3} 
\begin{figure}[!h]
		\includegraphics[width=\textwidth]{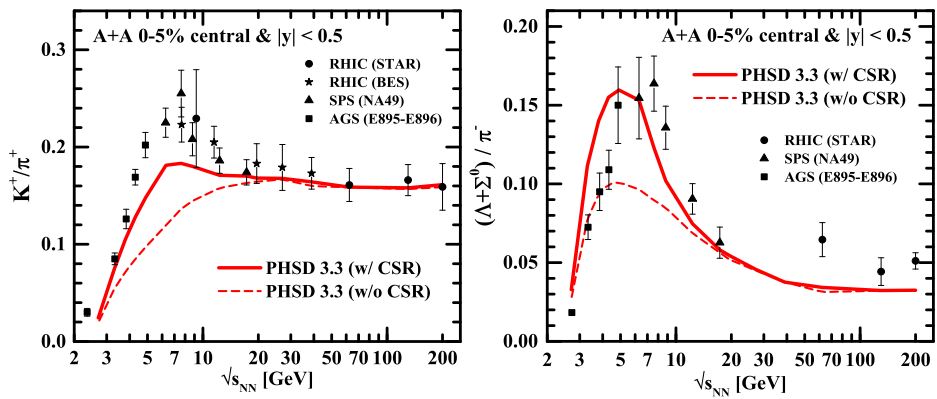}
	\caption{The ratios $K^+/\pi^+$ and $(\Lambda+\Sigma^0)/\pi^-$  at midrapidity
		from 5\% central A+A collisions as a function of the invariant
		energy $\sqrt{s_{NN}}$  in comparison to the experimental data from Ref.
		\cite{exp4}. The results from PHSD3.3 (with CSR) are displayed in
		terms of the full lines while the dashed lines show the results without CSR.}
	\label{fig6}
\end{figure}

Incorporating the effective masses (\ref{mss}) into the probability (\ref{schwinger}), we can determine the effects of CSR in the production of hadrons by string fragmentation.
In order to illustrate our findings we show the ratios $K^+/\pi^+$
and $(\Lambda+\Sigma^0)/\pi^-$ at midrapidity from 5\% central A+A
collisions in Fig. \ref{fig6} as a function of the invariant energy
$\sqrt{s_{NN}}$  in comparison to the experimental data available
\cite{exp4}. The solid (red) lines show the results from PHSD
(including CSR) while the dashed (red) line reflects the PHSD
results without CSR. It is clearly seen from Fig. \ref{fig6} that
the results in the conventional scenario (without
incorporating the CSR) clearly 
underestimate the ratios at low $\sqrt{s_{NN}}$ -- as found earlier
in Refs. \cite{Brat04,Weber} -- while the inclusion of CSR leads to
results significantly closer to the data. Especially, the rise of the
$K^+/\pi^+$ ratio at low invariant energy follows closely the experimental
excitation function when incorporating  'chiral symmetry
restoration'.

\section{Conclusions}

When comparing the results from the extended PHSD approach for the
ratios $K^+/\pi^+$ and $(\Lambda+\Sigma^0)/\pi^-$ from the different
scenarios we see in Fig. \ref{fig6} that the results from PHSD
fail to describe the data in
the conventional scenario \cite{Moreau:2015ika} without incorporating the CSR. 
Especially, the rise of the $K^+/\pi^+$ ratio at low
invariant energies follows closely the experimental excitation function when
including 'chiral symmetry restoration' in the string decay.
Nevertheless, the drop in this ratio again is due to 'deconfinement'
since there is no longer any hadronic string decay in a partonic
medium at higher energies. Accordingly, the experimental
'horn' in the excitation function is caused by chiral symmetry
restoration but also deconfinement is essential to observe a maximum
in the $K^+/\pi^+$  ratio.

\section*{Acknowledgements}
The authors acknowledge the support by BMBF, HIC for FAIR and the HGS-HIRe for FAIR.
The computational resources were provided by the LOEWE-CSC.

\end{document}